\documentstyle[12pt,german]{article}
\begin{document}
\hspace*{10cm}CERN-TH.7079/93\\
\vspace*{3cm}
\begin{center}
{\bf Constituent Quarks, Chiral Symmetry and the Nucleon
Spin}\footnote{Supported
in part by DFG-contract No. F412/19-1 and C.E.C. project SCI-CT-91-0729}$^, \,
$\footnote{Invited
talk given at the Leipzig Workshop ``Quantum Field Theory Aspects of High
Energy Physics,'' Kyffh"auser, Bad Frankenhausen, Germany, Sept. 20-24, 1993}\\
\bigskip
\bigskip
H. FRITZSCH\\
\bigskip
\bigskip
\it {Theory Division, CERN, Geneva}\\
\bigskip
\it and\\
\bigskip
\it {Sektion Physik, Universit"at M"unchen}\\
\bigskip
\bigskip
\bigskip
\bf {Abstract}\\
\end{center}
It is argued that the constituent quarks are expected to show a non-trivial
spin and flavor structure, due to the anomalous breaking of the chiral
symmetry in the U(1) sector.\\
\bigskip
\vspace*{5cm}
\begin{flushleft}
CERN.TH.7079/93\\
November 1993\\
\end{flushleft}
\newpage
\setcounter{page}{1}
\pagestyle{plain}
\noindent
{\Large Constituent Quarks, Chiral Symmetry and the Nucleon
Spin}\footnote[1]{Supported in part
by DFG contract No. F412/12-2 and C.E.C. project SCI-CT91-0729}$^, \, $
\footnote[2]{Invited talk given at the Leipzig Workshop ``Quantum Field Theory
Aspects of High Energy Physics,'' Kyffh"auser, Bad Frankenhausen, Germany,
Sept. 20/24, 1993}\\
\\
\\
Harald Fritzsch\\
\\
Theory Division, CERN, Geneva\\
and\\
Sektion Physik, Universit"at M"unchen\\
\\
\\
\\
\\
{\bf Abstract.} It is argued that the constituent quarks are expected to show
a non-trivial spin and flavor structure, due to the anomalous breaking of the
chiral symmetry in the U(1) sector.\\
\\
\\
\\
\\
Deep inelastic scattering reveals that the nucleon is a rather complicated
object consisting of an infinite number of quarks, antiquarks and gluons.
Although there is only scarce information about the internal structure of the
other strongly interacting particles, nobody doubts that the same is true for
all mesons and baryons. Nevertheless it seems that under certain circumstances
they behave as if they were composed of a single constituent quark and another
constituent antiquark or three constituent quarks. Examples are the magnetic
moments of the baryons, the spectroscopy of mesons and baryons, the
meson-baryon
couplings, the ratios of total cross sections like $\sigma (\pi N) \sigma (NN)$
and so on. Thus it seems to make sense to decompose the proton into three
pieces, into
three constituent quarks called U or D. A proton would have the composition
(UUD). The constituent quarks would carry the internal quantum numbers of the
nucleon.\\
\\
In deep inelastic scattering, one observes that a nucleon has the composition
$|u u d \bar q q...g...>$ (g: gluon, q = u,d,s), i.e., the quark density
functions
(which are scale dependent) are described by a valence quark and an essentially
infinite number of quark-antiquark pairs. One might be tempted to identify
the valence quark, defined by the corresponding quark density function, with
a constituent quark. This identification would imply that the three-quark
picture denoted above is nothing but a very rough approximation, and both
$\bar qq$-pairs and gluons need to be added to the picture. However, in this
case, one would not be able to understand why the model of a baryon consisting
of three constituent quarks works so well in many circumstances. It seems much
more likely to us that a constituent quark is a quasiparticle
which has a non-trivial internal structure on its own, i.e., consisting of a
valence quark, of many $\bar qq$-pairs and of gluons -- in short, it looks
like one third of a proton. Thus a constituent quark has an effective mass,
an internal size, and so on. This interpretation of a constituent quark is not
new; it was already
pointed out about 20 years ago$^{1}$. Nevertheless it is still unclear to what
extent it
can be derived from the basic laws of QCD, since it is deeply related to
non-perturbative aspects of QCD, like the confinement problem. In two
dimensions the constituent quarks can be identified with certain
soliton solutions of the QCD field equations$^2$.\\
\\
One way to gain deeper insights into the internal structure of the constituent
quarks is to consider spin problems, which is the topic of this talk. In the
constituent quark picture, it is, of course, assumed that the nucleon spin is
provided by the combination of the spins of the three constituent quarks. If
the latter have a non-trivial internal structure, the question arises whether
also the spin structure of the constituent quarks is a complex phenomenon, as
it
seems to be for the nucleon, or not.\\
\\
A simple model for the spin structure would be so assume that the spin of, say,
a constituent U-quark is provided by the valence u-quark inside it, and the
$\bar qq$-cloud and the gluonic cloud do not contribute to the spin. We shall
conclude that this ``na\"\i ve'' picture does not seem to be correct.\\
\\
Before we discuss the constituent quarks, let us summarize the results about
the spin structure of the proton. As usual we define the distribution functions
of the quarks of flavor q and helicity $+ 1/2 (-1/2)$ by $q_+ (q_-)$. The
lowest
moment of the structure function $g_1$, measured in the deep inelastic
scattering
of polarized leptons off hadronic targets, is given by the moments of these
quark densities $\Delta q$:\\
\begin{equation}
\int_0^1 g_1dx = \frac{1}{2}(\frac{4}{9} \Delta u + \frac{1}{9} \Delta d
+ \frac{1}{9} \Delta s),
\end{equation}
\begin{displaymath}
(\Delta u = \int_0^1 dx (u_+ +  u_- + \bar u_+ - \bar u_-)\, etc.).
\end{displaymath}
The spin density moments $\Delta q$ are determined by the nucleon matrix
elements of the associated axial-vector currents ($s_\mu$: spin vector):\\
\begin{equation}
\Delta q \cdot s_\mu = < p, s |\bar q \gamma_\mu \gamma_5 q|p, s>.
\end{equation}
The experimental data give$^2$:
\begin{equation}
\int^1_0 g_1dx = 0.114 \pm 0.012 \pm 0.026.
\end{equation}
The Bjorken sum rule which follows from the algebra of currents in QCD,
relates the difference of the $u/d$-moments to the axial-vector coupling
constant measured in $\beta$-deday:\\
\begin{equation}
\Delta u - \Delta d = g_A
\end{equation}
(we neglect radiative corrections of the order of $a_s / \pi $). For a
recent discussion of the experimental situation see ref. (4).
Using $SU_3$ one finds:\\
\begin{equation}
g_A = F + D, \,  \Delta u + \Delta d - 2 \Delta s = 3F - D.
\end{equation}
Here F and D are defined by the axial-vector matrix elements of the members
of the baryon octet. An analysis of the hyperon decays gives:$^5$

\begin{displaymath}
F = 0.47 \pm 0.04 \hspace{1cm}  D = 0.81 \pm 0.03
\end{displaymath}

\begin{equation}
\Delta u = 0.78 \pm 0.06 \hspace{1cm} \Delta d = -0.48 \pm 0.06
\end{equation}

\begin{displaymath}
\Delta s = -0.19 \pm 0.06
\end{displaymath}
\begin{displaymath}
\Delta \Sigma = \Delta u + \Delta d+ \Delta s = 0.10 \pm 0.17.
\end{displaymath}
An essential feature of the data is that the ``sea'' of the $\bar ss$-pairs in
the nucleon appears to be highly polarized; it contributes significantly to
the axial singlet charge. This implies that the ``Zweig'' rule does not seem
to work for the matrix elements of the axial baryonic current and can be
defined
as the axial baryon charge of the nucleon:
\begin{equation}
\Delta \Sigma \cdot s_\mu = <p| \bar u \gamma _\mu \gamma _5 u + \bar d \gamma
_\mu
\gamma _5 d + \bar s \gamma _\mu \gamma _5 s| p> = <p|j^{05}_\mu|p>.
\end{equation}
In a na\"\i ve wave function picture of the nucleon, the axial baryon number
correponds
to the portion of the nucleon spin carried by the quarks. Independent of a
specific wave function model, we can define $\Delta \Sigma $ as the relative
amount
of the nucleon spin carried by the intrinsic spins of the quarks. In the
simplest $SU_6$--type model of the nucleon, this quantity is one. In reality
it may depart significantly from one, due to the contributions of orbital
momenta and of the $\bar q q$-pairs or the gluons to the nucleon spin.
Nevertheless
it is surprising to observe that $\Delta \Sigma $ seems to be small compared
to one. However, we emphasize that the experiments give solely an information
about the axial baryonic charge of the nucleon and not about the spin. Only
in a non-relativistic $SU(6)$ type model, in which the quarks move in an
s-wave,
the axial baryonic charge and the spin of the nucleon, multiplied by two, are
both equal to one. There is no reason why $\Delta \Sigma $ could not be much
less than one, or even zero, if we doubt the validity of the ``na\"\i ve''
$SU(6)$
model.\\
\\
Of course, a possible vanishing of the axial-singlet nucleon charge must be
discussed in view of the fact that the octet axial charge are, of course,
different from zero. Nevertheless they depart substantially from the values
one obtains in a non-relativistic $SU(6)$ approach, which, for example,
predicts
$g_A / g_V = 5/3$, while in reality one has $g_A / g_V \approx 1.26$.\\
\\
Furthermore the octet charges obey the Goldberger-Treiman relations, which
relate the mass of the nucleon and the axial charges to the coupling and
decay constants of the pseudoscalar mesons. The latter act as massless
Nambu-Goldstone bosons in the chiral limit of QCD$^6$.
This suggests that also the value of the singlet axial charge is not unrelated
to the chiral symmetry of QCD and its dynamical breaking. For this reason it
is useful to examine the nucleon matrix element of the axial singlet current
in this respect. First we consider it in the chiral limit of $SU(3)_L \times
SU(3)_R$, in which $m_u = m_d =m_s = 0$. In this limit, the octet of axial
vector currents is conserved, while the singlet current is not conserved
due to the gluonic anomaly:
\begin{equation}
\partial^\mu j_\mu^{i5} = 0 (i = 1,2, \ldots 8)
\end{equation}

\begin{displaymath}
\partial^\mu j_\mu ^{05} = 3 \cdot \frac {^\alpha s}{2 \pi } \cdot tr G \tilde
G = a.
\end{displaymath}
It is known that this limit, in which the masses of the three light-quark
flavors are neglected, is not far away from the real world of hadrons. In the
limit, there exist eight massless pseudoscalar mesons, serving as Goldstone
bosons. However, the ninth pseudoscalar, the $\eta'$-meson, remains massive and
has a mass not far from its physical mass, i.e. about 900 MeV. The axial-vector
charges of the baryons are related to the coupling constants of the
pseudoscalar
mesons with the baryons by the Goldberger-Treiman relations, e.g., those for
the
pions ($f_\pi$: poin decay constant, M: nucleon mass):\\
\begin{equation}
2 M g_A = 2 f_\pi g_{\pi NN}.
\end{equation}
We remind the reader how these relations are obtained. The matrix element of
the axial-vector current in the octet channel can be described by two form
factors:\\
\begin{equation}
<p|j_\mu ^{i5}|p\,'> = \bar u(p) \left[G^i_1(q^2) \gamma_\mu \gamma_5 -
G^i_2 (q^2) q_\mu \gamma_5 \right] u(p\,') \hspace{2cm} q = p - p\,' , i = 1,2,
\ldots
8
\end{equation}
The induced pseudoscalar form factor $G_2 $ acquires a pole at $q^2 = 0$, since
the pion mass vanishes in the chiral limit:\\
\begin{equation}
G_2(q^2) = \frac{2f_\pi g_{\pi NN}}{q^2}.
\end{equation}
Due to the conservation of the current, one finds $2M \cdot G_1(0) = 2M g_A =
2f_\pi \cdot g_{\pi NN}$. We stress that this relation follows as the result
of an interplay between the axial-vector form factors $G_1$ and $G_2$. It is
the latter, which contains the pion pole. But the conservation of the current
leads to the constraint about $G_1$, i.e., to a condition about the axial
charge,
to the Goldberger-Treiman relation. In other words, the chiral symmetry allows
us to convert a statement about the divergence of the axial-vector current into
a statement about the matrix element of the current. Due to the pole in $G_2$
one finds a non-zero matrix element, even though the current is conserved. In
the absence of the pole, the chiral symmetry would be trivially fulfilled; the
nucleon mass would have to vanish.\\
\\
Let us consider the nucleon matrix element of the axial baryonic current in the
chiral limit:
\begin{equation}
<p|j^{05}_\mu|p \, '> = \bar u(p) (G^0_1 \gamma _\mu \gamma_5 - G^0_2 q_\mu
\gamma_5) u(p \, ').
\end{equation}
Here the induced pseudoscalar form factor does not have a Goldstone pole at
$q^2 = 0$. Instead of the Goldberger-Treiman relation, one finds after taking
the divergence and setting $q = 0^{7,8}$:\\
\begin{equation}
G^0_1 (0) = \Delta \Sigma = A(0)
\end{equation}
where A is the form factor of the anomalous divergence:\\
\begin{equation}
<p|3 \cdot \frac{^\alpha s}{2 \pi } tr G \tilde G|p \, ' > = 2 M A(q^2) \bar u
(p) i \gamma_5 u(p \, ').
\end{equation}
We conclude: In the chiral limit of vanishing quark masses, the axial baryonic
charge $\Delta \Sigma$ (``the spin of the nucleon'')
is nothing but the nucleon matrix element of the anomalous divergence, i.e.,
purely gluonic quantity. Not much is known about this quantity. Recently
one has succeeded to estimate $\Delta \Sigma $ by lattice simulations. One
finds $\Delta \Sigma $ to be significantly smaller than unity but not zero
($ \Delta \Sigma \approx 0.2$)$^{9, 10}$.\\
\\
It is interesting to note the fact that the singlet quantity $\Delta
\Sigma $ is a gluonic quantity while the octet spin densities, e.g., $\Delta u
+ \Delta d - 2 \Delta s$, are determined by the nucleon matrix elements of
quark bilinears, this indicates a substantial violation of the ``Zweig rule''
for the
axial-vector nonet. The latter would imply $\Delta s = 0$, and we would have
$\Delta \Sigma = \Delta u + \Delta d + \Delta s = \Delta u + \Delta d - 2\Delta
s$.
Thus the matrix
element of the anomalous divergence, a gluonic quantity, would have to be
equal to the matrix element of the eights component of the axial-vector octet.
There is no reason why this should be the case. If it were
true, the Ellis-Jaffe sum rule$^{11}$ could be used to calculate $\Delta \Sigma
$
in terms of $g_A$ and the D/F-ratio. The result ($\Delta \Sigma \approx 0.75$)
is in disagreement with the experimental results.\\
\\
We conclude: The violation of the ``Zweig rule'' in the pseudoscalar channel,
which is well known and caused by the QCD anomaly, implies via the mechanism
of spontaneous symmetry breaking another violation of this rule for the
nucleon-
matrix elements of the axial-vector current. The strength of this violation
is given by the magnitude of the spin density moment $\Delta s$. Therefore
it is not surprising that, in particular, this spin density moment appears
to be large.\\
\\
Apparently the violation of the ``Zweig rule'' is such that the axial singlet
charge
$\Delta \Sigma $ is rather small, perhaps even zero. Thus the constituent quark
model needs a revision which must take into account this effect, being a
consequence of the dynamics of chiral symmetry and its breaking. Below we shall
discuss such a revision, which is able to combine both chiral dynamics and the
``na\"\i ve'' constituent quark model$^{12, 13, 14}$.\\
\\
First we consider a simplified case, namely the one of QCD with the two flavors
u and d only. The strange quarks and the ``heavy'' flavors c, b and t are
disregarded. Furthermore, we assume $m_u = m_d = 0$, i.e., the chiral symmetry
$SU(2)_L x SU(2)_R$ is exactly fulfilled, and the pions are massless.\\
\\
Due to the QCD anomaly, the singlet pseudoscalar $\eta$ \,(quark composition
$(\bar u u + \bar d d) / \sqrt{2})$ has a mass of the order of the nucleon
mass $M$. The Goldberger-Treiman relation is exactly valid:
\begin{equation}
2M_ng_A = 2 F_\pi g_{\pi NN}.
\end{equation}
In the $SU(6)$-type constituent quark model, the axial-vector coupling constant
$g_A$ is given by the nucleon expectation value of the quark-spin operator
$\frac {1}{2} \sigma _z $:\\
\begin{equation}
g_A = < \sigma _z(u)> - < \sigma _z (d) > = 5/3
\end{equation}
where one has:\\
\begin{equation}
1/2 \sigma _z(u) = 2/3 \hspace{1cm} 1/2 \sigma _z(d) = - 1/6,
\end{equation}
\begin{displaymath}
1/2(\sigma _z(u) + \sigma _z(d) = 1/2 (= nucleon spin).
\end{displaymath}
In reality $g_A$ is not equal to 5/3, but about 1.26; i.e., the prediction of
the ``constituent model'' is violated by about 24 per cent. This violation can
be
understood without giving up the simple ideas of the constituent quark model,
as an effect due to orbital motions and relativistic effects. Thus in isovector
channel, both
the chiral dynamics and the constituent quark model do not contradict, but
rather supplement each other. This observation encourages us to consider the
``constituent quarks'' as separate entities. In a ``Gedankenexperiment,'' we
could consider a polarized ``constituent quark'' $Q(Q = U, D)$ and study its
coupling constants. They would also obey a Goldberger-Treiman type
relation$^{15} $:\\
\begin{equation}
2 M_q \tilde g_A = 2 F_\pi g_{\pi QQ}
\end{equation}
($M_q$: constituent quark mass, $\tilde g_A$ axial-vector coupling constant of
the
constituent quark, $g_{\pi QQ}$: pion-quark coupling constant).\\
\\
Suppose we consider the corresponding matrix elements of the vector and
axial-vector currents and relate them to the various moments of the quark
density functions. One finds na\"\i vely:
\begin{equation}
<U|\bar u \gamma _\mu \gamma _5 u|U> = p_\mu /M_U = \left(p_\mu / M_U \right)
\int^1_0 (u_+ + u_- - \bar u_+ - \bar u_-)dx
\end{equation}
\begin{displaymath}
<U|\bar d \gamma_\mu d|U> = 0
\end{displaymath}
\begin{displaymath}
<U|\bar u \gamma_\mu \gamma_5 u|U> = s_\mu \cdot \Delta u = s_\mu \cdot
\int^1_0
(u_+ + \bar u_+ - u_- - \bar u_-)dx = s_\mu \cdot 1
\end{displaymath}
\begin{displaymath}
<U|\bar d \gamma _\mu \gamma _5d|U> = 0
\end{displaymath}
\noindent
($s_\mu$: spin vector, $p_\mu$: four-momentum, the quark density functions
refer
to the U-quark and should carry an index $u$, which is not expicitly denoted
here.) These relations reflect the expectation that in a constituent U-quark
the quark density functions must be arranged such that the correct flavor
structure is obtained and that its total spin is carried by the u-flavor. The
d-flavor is not supposed to contribute to the spin.\\
\\
We could go further and be more specific about the structure of the quark
density functions. The success to the ``Zweig'' rule relies on the assumption
that $(\bar qq)$-pairs contribute little to the hadronic wave functions.
Correspondingly we can consider a limit in which the $(\bar qq)$-pairs are
neglected (``valence quark dominance''). In this limit we find for a U-quark:
\begin{equation}
\bar u_+ = \bar u_- = u_- =0
\end{equation}
\begin{displaymath}
d_+ = d_- = \bar d_+ = 0.
\end{displaymath}
Only the density function $u_+$ is different from zero. This is easily
understood if we consider the free quark model, in which the ``constituent
quarks'' and the ``current quarks'' are identical, and we have not only the
relations (20), but, in addition, the function $u_+$ is known: $u_+ = \delta
(x-1)$.\\
\\
Thus the essential difference between a ``constituent quark'' inside a hadron
and a free quark lies in the shape of the density functions $u_+$. The
confinement forces merely cause this function to depart from a
$\delta$-function
and to spread out over the available x-range.\\
\\
It turns out that the picture of a constituent quark described above is not
consistent with the constraints given by the chiral symmetry. In the
constituent
model, we have $\Delta d = 0$. This implies for a U-quark
that both the isoscalar and the isovector combinations of the spin density
moments are equal to one:
$\Delta u - \Delta d = \Delta u + \Delta d = 1$.\\
\\
The isovector part is determined
by the pion pole. If the isosinglet $\eta$-meson would also be a Goldstone
particle, the associated coupling constants would conspire such that the
isovector and isoscalar spin density moments would be equal, and the results of
the ``na\"\i ve'' constituent model would be obtained. However, due to the QCD
anomaly,
the isosinglet spin density function does not receive a Goldstone pole
contribution.
Instead it is given by the constituent quark matrix element of the anomalous
divergence:

\begin{displaymath}
\Delta u + \Delta d = A
\end{displaymath}

\begin{equation}
<U|2 \cdot \frac{^\alpha s}{2 \pi } tr G \tilde G| U> =
2 M_u A \bar u i \gamma _5u
\end{equation}

\begin{displaymath}
\Delta u = (1 + A) / 2 \hspace{1cm} \Delta d = (A-1) / 2.
\end{displaymath}
There is no dynamical reason why $A$ should be equal to one. If it were, the
spin
density moments would indeed reproduce the constituent quark model result. In
particular, $\Delta d$ would vanish. This is not ruled out a priori, but if it
were, it would be a miraculous coincidence.\\
\\
For all other values of $A$, the spin density moment $\Delta d$ does not
vanish.
We conclude that for $A \neq 1$ the constituent U quark must contain $(\bar
qq)$-
pairs. Thus a violation of the ``Zweig rule'' is automatically implied. It is
interesting that these pairs are generated by the same non-perturbative
mechanism
due to the gluon anomaly which causes the $\eta$-meson to acquire a mass and
not to act as a Goldstone particle in the chiral limit of $SU(2)_L \times
SU(2)_R$.\\
\\
Intuitively one can understand the violation of the ``Zweig rule'' discussed
above as follows. The chiral dynamics of a ``constituent quark'' would obey
the ``Zweig rule'' if it were surrounded by a cloud of $\pi$ and $\eta$
Goldstone
bosons. The Goldstone poles of the axial-vector current matrix elements would
imply,
via the Goldberger-Treimann relations in the isovector and isoscalar channel,
that the matrix elements obey the constraints given by the ``Zweig rule'' (in
particular
$\Delta d = 0$ for a U-quark and so on). However the QCD anomaly causes the
$\eta$-pole
at $q^2 = 0$ to disappear. As a result the ``Goldstone cloud'' of a U-quark
consists
only of $\pi$-mesons.\\
\\
Thus the dynamical structure of the constituent quark is drastically changed.
In partucular $(\bar uu)$ and $(\bar dd)$ pairs are generated, which modify the
spin structure.\\
\\
We note that these pairs cannot simply be regarded as the pairs inside virtual
$\pi$-mesons. Their presence is caused by the chiral dynamics, in particular,
by the Goldberger-Treimann relations for the axial-vector matrix elements.
Their
appearance is a non-perturbative phenomenon just like the generation of the
$\eta$-mass due to the gluonic anomaly.\\
\\
It is easy to apply similar considerations, as previously for the nucleon to
the constituent quarks. Also for them we should have $A \approx 0$, at least
to a good approximation. For $A = 0$ we find for a constituent U-quark that
the ``Zweig rule'' for the density moments is maximally violated:
\begin{equation}
\Delta u = 1/2 \hspace{1cm} \Delta d = -1/2.
\end{equation}
We can go further and specify the various density moments. If the ``Zweig
rule''
were valid (both $\pi $ and $\eta $ Goldstone modes present), we would have
\begin{equation}
\int u_+ dx =1, \hspace{0.5cm} u_- = \bar u_+ = \bar u_- = 0 \hspace{0.5cm}
d_+ = d_- = \bar d_+ = \bar d_- = 0.
\end{equation}
Such a constraint which is not invariant under the renormalization group can
only be imposed for a particular value of the energy scale $\mu $, which is
expected to be the characteristic hadronic energy scale. The removal of the
$\eta $ Goldstone pole causes a shift in the density moments, which we can
parametrize by two functions $h_+$ and $h_-$:
\begin{equation}
u_+ = u^v_+ + h_+ \hspace{1cm} \bar u_+ = d_+ = \bar d_+ = h_+
\end{equation}

\begin{displaymath}
u_- = \bar u_- = d_- = \bar d_- = h_-
\end{displaymath}
($u^v_+$: intrinsic density function of U-quark in the absence of the anomaly,
$\int u_+^vdx = 1$). We find for $A = 0$:
\begin{equation}
\Delta \Sigma = \Delta u + \Delta d = 1 + 4 \int ^1_0 (h_+ - h_-)dx = 0
\end{equation}
\begin{displaymath}
\Delta u - \Delta d = 1
\end{displaymath}
\begin{equation}
\int ^1_0 (h_+ - h_-) d_x = - 1/4.
\end{equation}
We observe that $\Delta \Sigma $ vanishes, because the constituent U-quark
contribution
to $\Delta \Sigma $ is cancelled by the pairs. A cancellation is only possible,
if the density function $h_-$ is different from zero. On the other hand
$h_+$ can be zero, in accordance with the sum rule (26). The simplest model
obeying the constraints discussed above is one in which we have
\begin{displaymath}
h_+ = 0, \hspace{2cm} \int h_-dx = 1/4,
\end{displaymath}
\begin{displaymath}
u_+ = u^v_+, \hspace{2cm} \int u_- dx = \int \bar u_-dx= \int d_-dx = \int
\bar d_-dx = 1/4
\end{displaymath}
\begin{equation}
d_+ = \bar d_+ = \bar u_+ = 0.
\end{equation}

\noindent
Thus we obtain in the case A = 0 the following picture of a polarized
constituent
U-quark in the $SU(2)_L \times SU(2)_R$ limit: The density function $u_+$,
which
describes the density of u-quarks polarized in the same direction as the
U-quark,
is unaffected by the QCD anomaly. The latter causes a large violation of the
``Zweig rule'' in the sense that $(\bar qq)$-pairs are generated. We shall
refer to this ``cloud'' of $(\bar qq)$-pairs as the ``anomaly cloud''. The
density functions $u_-, \bar u_-, d_-, \bar d_-$
are different from zero; i.e., the pairs are polarized opposite to the original
constituent quark. The sum of all (anti)-quark spins is zero. Thus for $A = 0$
the quarks do not contribute to the spin of the constituent quark. The latter
is provided either by the orbital angular momentum of the pairs or by gluons or
both. This can be seen as
follows. If we would turn off the QCD anomaly (e. g. formally by setting
$n_c = \infty$), the ``na\"\i ve'' picture should hold, i.e., the spin of the
U-quark
is carried by the valence quark $u_v$. Once the anomaly is introduced, the
u-valence quark continues to contribute its spin, but the ($\bar qq$) pairs
cancel the latter. Their total angular momentum must be zero. Otherwise, the
introduction of the anomaly would violate the conservation of angular
momentum.\\
\\
Thus we have:
\begin{equation}
J_z(U) = + 1/2 = J_z(u_v) + J_z(cloud) + (L_z(cloud) + L_z(gluons))
\end{equation}
\begin{displaymath}
= + 1/2 + (-1/2) + (+ 1/2).
\end{displaymath}
\noindent
In the case $A \neq 0$, the cancellation between the spin of the valence quark
and the spins of the ``anomaly cloud'' and of the gluons would not be
complete, but the sum of the spins and of the orbital angular momenta of the
pairs in the ``anomaly cloud'' would still be zero.\\
\\
Finally we consider the case of the three-light flavors u,d,s. In the chiral
limit of $SU(3)_L \times SU(3)_R$, we obtain for a constituent U-quark in
analogy
to eq. (27):
\begin{equation}
\Delta U = 2/3 \hspace{1.5cm} \Delta d = - 1/3 \hspace{1.5cm} \Delta s = -1/3.
\end{equation}
In the symmetry limit, the ``anomaly cloud'' is, of course, $SU(3)$ symmetric.
In reality symmetry breaking will be present. The result will be that the
effects of the $(\bar ss)$ pairs are somewhat reduced compared to those of the
$(\bar uu)$ and $(\bar dd)$ pairs. For example, in a U-quark we expect:
$|\Delta d| >| \Delta s|$. The actual spin density momenta of the U, D
constituent quarks
 will lie between the extreme case of $SU(2) \times SU(2)\, \, (\Delta d = -
1/2$ for a
 U-quark) and of $SU(3) \times SU(3)\, \, (\Delta d = - 1/3$ for a U-quark).
\\
\\
It has been argued that the anomaly could contribute to the axial-singlet
 charge if gluons are highly polarized nucleons. In this case their
contribution
 to the singlet charge could be calculated perturbatively $^{16,17}$. In our
 approach we see no reason for a large gluonic polarization. Thus the effect
 discussed in ref. (16,17) would be negligible in comparison to the
non-perturbative
 phenomenon discussed here.\\
\\
The smallness of the axial-singlet charge, parametrized above by the parameter
 A, follows also within the Skyrme-type model, as discussed in ref. (18). The
 connection of this model to the scheme discussed here remains unclear,
althouth
 some common features exist. In our approach we would also expect that in the
 case of one flavor the spin of a constituent quark is cancelled partially or
 fully by the ``anomaly cloud.'' Thus we see no qualitative difference between
 the cases of one or two (three, $\ldots$) flavors. On the other hand in the
 Skyrme model the case of one flavor is not defined.\\
\\
The picture of ``constituent quarks'' carrying a polarized ``anomaly cloud''
 described here, implies that many aspects of hadronic physics, especially
those in
 which polarization and spin aspects are relevant, must be reconsidered. Among
 them are the magnetic moments of the baryons, the polarization phenomena of
 hyperons in hadronic processes and the spin asymmetries observed in the strong
 interaction processes. Many further tests of the ideas presented here can be
 envisaged, once spin asymmetries can be measured in electroweak lepton-hadron
 reactions at high energies. The generation of a cloud of ($\bar qq$)-pairs
 by the QCD anomaly reminds us of the ``Cooper pairs'' in the BCS-theory of
 superconductivity. Indeed there are some analogies between superconductivity
 and hadronic physics in the chiral limit, e.g., the appearance of the mass
gap,
 which in QCD is related to the anomaly as well as to the dynamical breaking
 of scale invariance and the chiral symmetry, and the presence of pairing
forces,
 which in QCD are responsible for the removal of the Goldstone pole in the
 singlet axial-vector channel.\\
 \\
 According to standard meson dominance ideas, the axial-singlet charge of the
 nucleon is related to the coupling constants of the neutral $1^{+-}$
-axial-vector
 mesons. Since apparently the axial-singlet charge is influenced strongly by
 the gluon anomaly, we would expect that the coupling constants of the
 axial-vector mesons and their mass and mixing pattern are also influenced
 by the gluon anomaly. In particular the mass eigenstates of the axial-vector
 mesons would show a quark composition similar to the pseudoscalar mesons
 (where the gluon anomaly plays an essential role) and not similar to the
 vector mesons (where the mixing between $\bar u u, \, \bar dd$ and $\bar ss$
is
 very small)$^{19}$.\\
\\
In this lecture, I have described why polarized constituent quarks should be
 surrounded by a cloud of polarized quark-antiquark pairs. Our reasoning was
 entirely based on phenomenological arguments. It would be interesting to see
 how these polarized pairs are generated dynamically, via those
non-perturbative
 effects, due to instantons and the like, which are responsible also for the
QCD mass
 gap and the breaking of the chiral symmetry in the axial-singlet channel. An
 explicit dynamical model along these lines is not yet available.\\
\\
\newpage
\noindent
{\bf Acknowledgements}\\
\\
It is a pleasure to thank the members of the Leipzig group, especially
Prof. B. Geyer and Dr. M. Ilgenfritz, for their great efforts to organize
this meeting in these colourful mountains, which were at the beginning of the
Middle Age one of the centres of Western Europe culture.\\
\newpage
\noindent
{\bf References}
\begin{tabbing}
10.\quad\= \kill
\renewcommand{\baselinestretch}{1}
1.\>H. Fritzsch and M. Gell-Mann, Proc. of XVI. International Conference on\\
\>High Energy Physics, Vol. 2, p. 135 (Chicago, 1972).\\
\\
\enspace 2.\>J. Ellis, Y. Frishman, A. Hanany and M. Karliner, Nucl. Phys. B382
(1992) 189.\\
\\
\enspace 3.\>G. Baum et al., Phys. Lett. 51 (1983 1135.\\
\>J. Ashman et al., Phys. Lett. B206 (1988) 364.\\
\>V.W. Hughes et al., Phys. Lett. B212 (1988) 511.\\
\\
\enspace 4.\>J. Ellis and M. Karliner, {\it Spin Structure Functions preprint
CERN-TH.7022/93}.\\
\\
\enspace 5.\>M. Bourquin et al., Zeitschrift f"ur Physik C21 (1983) 27.\\
\\
\enspace 6.\>See H. Pagels, Phys. Rep. 16 (1975) 219.\\
\\
\enspace 7.\>H. Fritzsch, Phys. Lett. 229B (1989) 1605; Ibid., 122.\\
\\
\enspace 8.\>G. Venziano, Mod. Phys. Lett. A17 (1989) 1605.\\
\\
\enspace 9.\>J.E. Mandula and M.C. Ogilvie, preprint HEP LAT / 9208009.\\
\\
10.\>R. Altmeyer et al., preprint DESY-92-187 (1992).\\
\\
11.\>J. Ellis and R.L. Jaffe, Phys. Rev. D9 (1974) 1444; D10 (1974) 1669.\\
\\
12.\>H. Fritzsch, Phys. Lett. 256B (1991) 75.\\
\\
13.\>H. Fritzsch, Phys. Lett. A5 (1990) 625.\\
\\
14.\>U. Ellwanger und B. Stech, Phys. Lett. B241 (1990) 449; Z. Physik C 49
(1991) 683.\\
\\
15.\>M. Goldberger and S.B. Treiman, Phys. Lett. B212 (1988) 391.\\
\\
16.\>G. Altarelli and G. Roos, Phys. Lett. B212 (1988) 391.\\
\\
17.\>R. Carlitz, J. Collins and A. Mueller, Phys. Lett. B214 (1988) 229.\\
\\
18.\>S. Brodsky, J. Ellis and M. Karliner, Phys. Lett. B206 (1988) 309.\\
\\
19.\>M. Birkel and H. Fritzsch, in preparation.\\
\end{tabbing}
\end{document}